\documentclass[twocolumn,showpacs,preprintnumbers,aps,prl,floatfix]{revtex4}
\usepackage{graphicx}
\usepackage{dcolumn}
\usepackage{amsfonts}
\usepackage{latexsym}
\usepackage{amssymb} 
\usepackage{verbatim}
\usepackage{amsthm}
\usepackage{amsmath}

\begin{document}
%
\title{
  Torsion bar antenna in the proper reference frame with rotation
}
\author{
  Kouji Nakamura${}^{1}$ \footnote{Electronic address: kouji.nakamura@nao.ac.jp}
  and
  Masaki Ando${}^{1,2}$
}
\affiliation{%
  ${}^{1}$TAMA Project Office, Optical and Infrared Astronomy Division,
  National Astronomical Observatory,\\
  Osawa 2-21-1, Mitaka, Tokyo 181-8588, Japan\\
  ${}^{2}$Department of Physics, The University of Tokyo,
  Hongo 7-3-1, Tokyo 113-0033, Japan
}
\date{\today}
\begin{abstract}
  The resultant response of the rotating torsion bar antenna for
  gravitational waves discussed in [M.~Ando et al.,
  Phys. Rev. Lett. {\bf 105} (2010), 161101.] is re-investigate
  from a general-relativistic point of view.
  To do this, the equation of motion of a free falling particle
  in the proper reference frame of a rotating observer is used.
  As a result, the resultant response derived in the above paper
  is also valid even when $\omega_{g}\sim\Omega$, where
  $\omega_{g}$ and $\Omega$ are the angular frequencies of
  gravitational waves and the rotation of the antenna,
  respectively.
\end{abstract}
\pacs{04.30.-w, 04.80.Nn, 95.55.Ym}
\maketitle




TOrsion-Bar Antenna (TOBA) is one of novel types of
gravitational-wave antenna for low-frequency observations.
This antenna is formed by two bar-shaped test masses, arranged
parallel to the $x-y$ plane and orthogonal to each other.
Each bar is supported at its center, so as to rotate around the
$z$ axis.
When gravitational waves pass through this antenna, tidal forces
by the gravitational waves will appear as differential angular
changes in these bars.
These changes are extracted as a gravitational-wave signal by
using a sensitive sensor, such as a laser interferometer.
A characteristic feature of this antenna is that it can expand
the observation band to lower frequencies by using modulation
and up-conversion of gravitational-wave signals by rotation of
the antenna.


A similar concept was proposed more than 40 years ago
as a heterodyne detector for circular-polarized
gravitational-waves~\cite{V.B.Braginsky-V.S.Nararenko-1971,C.W.Misner-T.S.Thorne-J.A.Wheeler-1973}.
Recently, the idea for the up-conversion of low-frequency
gravitational waves was re-investigated~\cite{M.Ando-etal-2010}
and a space-borne prototype antenna was
operated~\cite{W.Kokuyama-2012}.
In Ref.~\cite{M.Ando-etal-2010}, it was considered the situation
where the frequency of gravitational waves is much smaller than
the rotation of the antenna.
In this situation, the tidal force due to gravitational waves is
almost stationary and the antenna rotates in this stationary tidal
force field.
Based on this intuitive picture, the response of the antenna was
derived as
\begin{eqnarray}
  \label{eq:resultant-output-Ando}
  \ddot{\theta}_{diff}
  = 
  \alpha \left[
    \ddot{h}_{\times} \cos(2\Omega t)
    +
    \ddot{h}_{+} \sin(2\Omega t)
  \right]
  ,
\end{eqnarray}
where $\theta_{diff}$ is the resultant output of the antenna,
$\alpha$ is the shape factor of the antenna, $h_{\times}$ and
$h_{+}$ are the two independent polarization components of
gravitational waves propagate along $z$ axis, and $\Omega$ is
the angular velocity of the rotation of the antenna.
From this equation, it was concluded that the
gravitational-waves signal modulated by the rotation; a
gravitational-wave signal with an angular frequency 
of $\omega_{g}$ is up- and down-converted to appear at
$\omega_{g}\pm 2\Omega$ frequencies.
However, due to the above intuitive picture, the resultant
output (\ref{eq:resultant-output-Ando}) was valid only when
$\omega_{g}\ll\Omega$.


The purpose of this paper is re-derivation of the resultant
output (\ref{eq:resultant-output-Ando}) of the rotating TOBA
from a general-relativistic point of view.
Usually, the geodesic deviation equation is the basic equation
to estimate the force which affects to gravitational-wave
detectors.
However, in the case of the response of the rotating TOBA, we
cannot apply the geodesic deviation equation, because the world
line of the test mass in the rotating TOBA is not geodesic.
Therefore, in this article, we estimate the torque, which
affects to the rotating TOBA test mass, through the proper
reference frame for a rotating observer.


The proper reference frame for an accelerating and rotating
observer was discussed in the text 
book~\cite{C.W.Misner-T.S.Thorne-J.A.Wheeler-1973}. 
After the publication of this text book, Ni and
Zimmermann~\cite{W.-T.Ni-M.Zimmermann-1978} derived the metric
which is accurate to the second order in proper distance from
the origin of coordinates.
They also derived the equations of motion for freely falling
particles, which is accurate to first order.
Their equation of motion contains many terms which represent
many type of effects of inertial forces and forces due to the
Riemann curvature.


The situation discussed by Ni and
Zimmermann~\cite{W.-T.Ni-M.Zimmermann-1978} is appropriate to
the rotating TOBA.
Therefore, we use the proper reference frame with the rotation
discussed by Ni and Zimmermann~\cite{W.-T.Ni-M.Zimmermann-1978}
to estimate the gravitational-wave torque which affects to the
rotating TOBA response.
The metric and the equation of motion discussed in this paper is
the special case of the proper reference frame discussed by Ni
and Zimmermann~\cite{W.-T.Ni-M.Zimmermann-1978}.
Through these metric, we show the resultant output
(\ref{eq:resultant-output-Ando}) is valid even if
$\omega_{g}\sim\Omega$.
Throughout this paper, we use the natural unit in which the
light velocity is unity.






Here, we consider the situation where the TOBA is rotating
around $Z$ axis, where the center of the TOBA is setting
$X=Y=Z=0$.
Consider the world line $P_{0}(\tau)$ of the center of TOBA with
four-velocity $u^{a}(\tau)$ and four-rotation $\omega^{a}$ in
gravitational waves with Riemann tensor $R_{abc}^{\;\;\;\;\;\;d}$.
The orthogonal tetrad $e^{a}_{(\alpha)}$, which observers at the
center of the TOBA carries, transports according
to~\cite{C.W.Misner-T.S.Thorne-J.A.Wheeler-1973}
\begin{eqnarray}
  \label{eq:W.-T.Ni-M.Zimmermann-1978-kouchan-1}
  u^{b}\nabla_{b}e^{a}_{(\alpha)} = - \Omega^{ab} e_{b(\alpha)},
\end{eqnarray}
where
\begin{eqnarray}
  \label{eq:W.-T.Ni-M.Zimmermann-1978-kouchan-2}
  \Omega^{ab} &:=& a^{a}u^{b} -  a^{b}u^{a} + \epsilon^{abcd} u_{c} \omega_{d}
  , \\
  \label{eq:W.-T.Ni-M.Zimmermann-1978-kouchan-3}
  a^{a}(\tau) &:=& u^{b}\nabla_{b}u^{a}.
\end{eqnarray}
Even when the TOBA is rotating, the world line $P_{0}(\tau)$ of
the center of TOBA is still a geodesic with the four-velocity
$u^{a}$ if the laboratory can be regarded as a local inertial
system.
Furthermore, we concentrate only on the case where the angular
velocity of rotation is constant.
Therefore, in this paper, we may regard that
\begin{eqnarray}
  \label{eq:W.-T.Ni-M.Zimmermann-1978-kouchan-4}
  a^{a}(\tau) = 0, \quad u^{b}\nabla_{b}\omega^{a}(\tau) = 0,
\end{eqnarray}
along $P_{(0)}(\tau)$.


At any event $P_{0}(\tau)$, we consider geodesics
$P(\tau;n^{a};s)$ orthogonal to $u^{a}(\tau)$, where $n^{a}$ is
the unit vector tangent to a particular geodesic at
$P_{0}(\tau)$, and $n^{a}u_{a}=0$.
At each event $P_{0}(\tau)$, a proper distance $s$ along any
geodesic $P(\tau;n^{a};s)$ with tangent vector $n^{a}$ is
assigned by the local coordinates
\begin{eqnarray}
  \label{eq:x0-def}
  X^{\hat{0}} &:=& \tau,\\
  \label{eq:xhatj-def}
  X^{\hat{j}} &:=& s n^{a} e^{(\hat{j})}_{a} = s \alpha^{\hat{j}},
\end{eqnarray}
where $\alpha^{\hat{j}}$ is the spatial direction cosine and
$\alpha^{\hat{j}} = X^{\hat{j}}/s$ with
$s^{2}=(X^{\hat{i}})^{2}=(X^{\hat{1}})^{2}+(X^{\hat{2}})^{2}+(X^{\hat{3}})^{2}$,
and $X^{\hat{0}}=\tau=T$, $X^{\hat{1}}=X$,  $X^{\hat{2}}=Y$, and
$X^{\hat{3}}=Z$.
This means $n^{a} = \alpha^{\hat{j}} e^{a}_{(\hat{j})}$.
This coordinate system is well-defined for events near the world
line $P_{0}(\tau)$ if the ``light-cylinder'' has not been
reached ($s\ll 1/|\Omega|$), nor curvature has not yet caused
geodesics to cross ($s\ll 1/|R_{abc}^{\;\;\;\;\;\;d}|^{1/2}$),
and the Riemann tensor has not yet changed much from its value
on the world
($s\ll|R_{abc}^{\;\;\;\;\;\;d}|/|\partial_{a}R_{bcd}^{\;\;\;\;\;\;e}|$).
In the case of gravitational waves, the last condition
corresponds to the fact $s$ should be much smaller than the
wavelength of gravitational waves.

Using the above coordinate system $\{X^{\hat{\mu}}\}$, Ni and
Zimmermann~\cite{W.-T.Ni-M.Zimmermann-1978} derived the
second-order expansion of the metric near $P_{0}(\tau)$ as
\begin{eqnarray}
  ds^{2}
  &=&
  - (dX^{\hat{0}})^{2} \left[
    1
    + (\omega^{\hat{l}}X^{\hat{l}})^{2}
    - (\omega)^{2}X^{\hat{l}}X^{\hat{l}}
  \right.
  \nonumber\\
  && \quad\quad\quad\quad\quad\quad
  \left.
    + R_{\hat{0}\hat{l}\hat{0}\hat{m}}X^{\hat{l}}X^{\hat{m}}
  \right]
  \nonumber\\
  &&
  +
  2 dX^{\hat{0}} dX^{\hat{i}} \left(
    \epsilon_{\hat{i}\hat{j}\hat{k}} \omega^{\hat{j}} X^{\hat{k}}
    -
    \frac{2}{3} R_{\hat{0}\hat{l}\hat{i}\hat{m}} X^{\hat{l}}X^{\hat{m}}
  \right)
  \nonumber\\
  &&
  +
  dX^{\hat{i}}dX^{\hat{j}} \left(
    \delta_{\hat{i}\hat{j}}
    -
    \frac{1}{3} R_{\hat{i}\hat{l}\hat{j}\hat{m}} X^{\hat{l}} X^{\hat{m}}
  \right)
  \nonumber\\
  &&
  +
  O(dX^{\hat{\mu}}dX^{\hat{\nu}}X^{\hat{l}}X^{\hat{m}}X^{\hat{k}}),
  \label{eq:W.-T.Ni-M.Zimmermann-1978-18-again}
\end{eqnarray}
where $\omega^{\hat{l}}$ and
$R_{\hat{\alpha}\hat{\beta}\hat{\mu}\hat{\nu}}$ are evaluated on
the world line $P(\tau)$ at time $X^{\hat{0}}=\tau$.


To calculate the coordinate acceleration of a freely falling
body, we use the geodesic equation in the form
\begin{eqnarray}
  \label{eq:W.-T.Ni-M.Zimmermann-1978-19-again}
  \frac{d^{2}X^{\hat{i}}}{(dX^{\hat{0}})^{2}}
  +
  \left(
    \Gamma^{\hat{i}}_{\;\;\;\hat{\mu}\hat{\nu}}
    -
    \Gamma^{\hat{0}}_{\;\;\;\hat{\mu}\hat{\nu}}
    \frac{dX^{\hat{i}}}{dX^{\hat{0}}}
  \right)
  \frac{dX^{\hat{\mu}}}{dX^{\hat{0}}}
  \frac{dX^{\hat{\nu}}}{dX^{\hat{0}}}
  =
  0
\end{eqnarray}
and substitute into it the first-order expansion of the
$\Gamma$'s.
Defining $W^{i}:=dX^{\hat{i}}/dX^{\hat{0}}$, the velocity
measured by the accelerated rotating observer, the resulting
coordinate acceleration is
\begin{eqnarray}
  \frac{d^{2}X^{\hat{i}}}{(dX^{\hat{0}})^{2}}
  &=&
  -
  R_{\hat{0}\hat{j}\hat{0}}^{\;\;\;\;\;\;\hat{i}} X^{\hat{j}}
  \nonumber\\
  &&
  - \left(\vec{\omega}\times(\vec{\omega}\times\vec{X})\right)^{i}
  - 2 \left(\vec{\omega}\times\vec{W}\right)^{i}
  \nonumber\\
  &&
  - 2 R_{\hat{0}\hat{k}\hat{j}}^{\;\;\;\;\;\;\hat{i}} X^{\hat{k}} W^{\hat{j}}
  + 2 R_{\hat{0}\hat{l}\hat{j}}^{\;\;\;\;\;\;\hat{0}} X^{\hat{l}} W^{\hat{j}} W^{\hat{i}}
  \nonumber\\
  &&
  +
  \frac{2}{3}
  \bar{R}_{\hat{m}\hat{k}\hat{j}}^{\;\;\;\;\;\;\hat{i}}
  X^{\hat{m}}
  W^{\hat{j}}
  W^{\hat{k}}
  -
  \frac{2}{3}
  \bar{R}_{\hat{l}\hat{k}\hat{j}}^{\;\;\;\;\;\;\hat{0}}
  X^{\hat{l}}
  W^{\hat{j}}
  W^{\hat{k}}
  W^{\hat{i}}
  \nonumber\\
  &&
  +
  O((X^{\hat{i}})^{2})
  .
  \label{eq:W.-T.Ni-M.Zimmermann-1978-20-TOBA}
\end{eqnarray}
The first line in
Eq.~(\ref{eq:W.-T.Ni-M.Zimmermann-1978-20-TOBA}) is the tidal
force due to the curvature near $P_{0}(\tau)$.
The first term in the second line of
Eq.~(\ref{eq:W.-T.Ni-M.Zimmermann-1978-20-TOBA}) is the
centripetal force and the second term is the Coriolis
force.
These terms should be neglected in the case of TOBA's
configuration as discussed below.




Here, we consider the configuration of the rotating TOBA.
TOBA essentially measures the relative rotation of the two bars
which induced by the tidal force of gravitational waves.
To measure the tidal force due to gravitational waves, the
center of bars are fixed at the point $X=Y=0$ but free for
rotational modes.
The test mass is aligned with in $(X-Y)$-plane and the rotation
axis of TOBA is chosen so that $\omega^{a}=\Omega e_{(Z)}^{a}$.
The shape of the test mass is symmetric which means the density
distribution $\rho$ is symmetric in all three axes.


In this configuration, the direction of the centripetal force
$-\left(\vec{\omega}\times(\vec{\omega}\times\vec{X})\right)^{i}$,
which is the first term in the second line of
Eq.~(\ref{eq:W.-T.Ni-M.Zimmermann-1978-20-TOBA}), is the
direction along the bar.
This force canceles out if we sum the this force along the
bar.
Similarly, the velocity $W^{i}$ of the bar is restricted to the
only rotational motion around the $e_{(Z)}^{a}$ axis.
In this case, the Coriolis force
$-2\left(\vec{\omega}\times\vec{W}\right)^{i}$, which is the 
second term in the second line of
Eq.~(\ref{eq:W.-T.Ni-M.Zimmermann-1978-20-TOBA}), also directs to
the direction along the bar.
This force also cancels out when we sum this force along the
bar.
Therefore, we may neglect the second line in
Eq.~(\ref{eq:W.-T.Ni-M.Zimmermann-1978-20-TOBA}).
Furthermore, we regard that the velocity $W^{i}$ is induced by
gravitational waves, i.e., $W^{i}=O(h)$.
As far as we concentrate only on the linear effect of
gravitational waves, we may neglect the third and the fourth
lines in Eq.~(\ref{eq:W.-T.Ni-M.Zimmermann-1978-20-TOBA}).


In the case where the mass distributes as the mass density
$\rho$, the force $F_{i}$ induced by GW on an volume element
$dV$ of TOBA's test mass is given by
\begin{eqnarray}
  F_{\hat{i}}dV = - \rho R_{\hat{0}\hat{j}\hat{0}\hat{i}} X^{\hat{j}} dV
\end{eqnarray}
This force is also derived from the potential 
\begin{eqnarray}
  U
  &:=&
  - \int dV \int dX^{\hat{i}} F_{\hat{i}},
  \nonumber\\
  &=&
  \frac{1}{2} R_{\hat{0}\hat{j}\hat{0}\hat{i}} \int dV \rho X^{\hat{i}}X^{\hat{j}}
  .
\end{eqnarray}
The torque $F_{gw}$ induced by the gravitational wave is given
by
\begin{eqnarray}
  F_{gw}
  &=&
  - \frac{\partial U}{\partial\theta}
  =:
  - \frac{1}{2} R_{\hat{0}\hat{j}\hat{0}\hat{i}}
  q^{\hat{i}\hat{j}}
  ,
  \label{eq:Forces_affects_the_rotating_TOBA}
\end{eqnarray}
where $q^{\hat{i}\hat{j}}$ is the dynamic quadruple moment
tensor~\cite{H.Hirakawa-K.Narihara-M.-K.Fujimoto-1976}.
For bar rotation $q^{XX}=-q^{YY}=-\int\rho(2XY)dV$ and
$q^{XY}=q^{YX}=\int\rho(X^{2}-Y^{2})dV$.




To evaluate curvature components in
Eq.~(\ref{eq:Forces_affects_the_rotating_TOBA}), we consider the
gravitational wave solution with a flat spacetime background
$g_{ab}=\eta_{ab}+h_{ab}$, where $h_{ab}$ is transverse
traceless, i.e., $\eta^{ab}h_{ab}=0=\eta^{ad}\partial_{d}h_{ab}$.
In a inertia frame, the background metric is given by
$\eta_{ab}$ $=$ $-(dt)_{a}(dt)_{b}$ $+(dx)_{a}(dx)_{b}$
$+(dy)_{a}(dy)_{b}$ $+(dz)_{a}(dz)_{b}$, and we assume the
gravitational wave propagates along the $z$-axis:
\begin{eqnarray}
  h_{ab}
  &=&
  h_{+}(t+z)\left(
    (dx)_{a}(dx)_{b} - (dy)_{a}(dy)_{b}
  \right)
  \nonumber\\
  &&
  +
  2 h_{\times}(t+z) (dx)_{(a}(dy)_{b)}
  .
\end{eqnarray}


Since we consider the rotating TOBA with the rotating axis $z$, 
this rotational frame $\{T,X,Y,Z\}$ is given by
\begin{eqnarray}
  && T = t, \\
  && X = x \cos\Omega t - y \sin\Omega t, \\
  && Y = x \sin\Omega t + y \cos\Omega t, \\
  && Z = z.
\end{eqnarray}
From this coordinate transformation, the flat metric $\eta_{ab}$
and TT-gauge gravitational wave $h_{ab}$ are given
by~\cite{KNcomment-gauge}:
\begin{eqnarray}
  \eta_{ab}
  &=&
  -
  \left(
    1
    - \Omega^{2} \left(Y^{2} + X^{2}\right)
  \right)
  (dT)_{a} (dT)_{b}
  \nonumber\\
  &&
  + 2 \Omega (dT)_{(a} \left( Y dX - X dY \right)_{b)}
  \nonumber\\
  &&
  + (dX)_{a}(dX)_{b}
  + (dY)_{a}(dY)_{b}
  + (dZ)_{a}(dZ)_{b}
  \nonumber\\
  \label{eq:flat-metric-in-rotation}
\end{eqnarray}
and
\begin{widetext}
\begin{eqnarray}
  h_{ab}
  &=&
  \Omega^{2} \left(
    h_{+} \left(
      \cos(2\Omega T) (Y^{2} - X^{2}) - 2 \sin(2\Omega T) XY
    \right)
    + h_{\times} \left(
      \sin(2\Omega T) (X^{2} - Y^{2}) - 2 \cos(2\Omega T)) XY
    \right)
  \right) (dT)_{a}(dT)_{b}
  \nonumber\\
  &&
  + 2 \Omega \left(
    h_{+} \left(
      \cos(2\Omega T) Y - \sin(2\Omega T) X
    \right)
    -
    h_{\times} \left(
      \sin(2\Omega T) Y + \cos(2\Omega T)) X
    \right) 
  \right) (dX)_{(a}(dT)_{b)}
  \nonumber\\
  &&
  + 2 \Omega \left(
    h_{+} \left(
      \cos(2\Omega T) X + \sin(2\Omega T) Y
    \right) 
    +
    h_{\times} 
    \left(
      \cos(2\Omega T)) Y - \sin(2\Omega T) X
    \right)
  \right) (dY)_{(a}(dT)_{b)}
  \nonumber\\
  &&
  +
  \left(
    h_{+} \cos(2\Omega T) - h_{\times} \sin(2\Omega T) 
  \right)
  \left(
    (dX)_{a}(dX)_{b} - (dY)_{a}(dY)_{b}
  \right)
  \nonumber\\
  &&
  + 2 \left(
    h_{+} \sin(2\Omega T) + h_{\times} \cos(2\Omega T))
  \right) (dX)_{(a}(dY)_{b)}
  \label{eq:z-propagate-TT-GW}
\end{eqnarray}
\end{widetext}
We note that the metric (\ref{eq:flat-metric-in-rotation}) is
consistent with Eq.~(\ref{eq:W.-T.Ni-M.Zimmermann-1978-18-again}).


In this rotational coordinate system, the components of the
Riemann curvature which are necessary for the evaluation of
Eq.~(\ref{eq:Forces_affects_the_rotating_TOBA}) are summarized
as
\begin{eqnarray}
  R_{TXTX}
  &=& 
  \frac{1}{2} \left(
    \sin (2 \Omega T) \ddot{h}_{\times}
    -
    \cos (2 \Omega T) \ddot{h}_{+}\right)
  \nonumber\\
  &=&
  - R_{TYTY}
  , \\
  R_{TXTY}
  &=& 
  -\frac{1}{2} \left(
    \cos (2 \Omega T) \ddot{h}_{\times}
    +
    \sin (2 \Omega T) \ddot{h}_{+}
  \right)
  .
\end{eqnarray}


Since $q^{XX}=q^{YY}=0$ and $q^{XY}=:I\alpha$ in the case of the
thin bar aligned along $X$ axis, the torque which affects this
thin bar is given by
\begin{eqnarray}
  F_{gw(xbar)}
  &=&
  - \frac{I\alpha}{2}
  \left(
    \cos(2\Omega T)\ddot{h}_{\times}
    +
    \sin(2\Omega T)\ddot{h}_{+}
  \right)
  ,
  \label{eq:Forces_affects_the_rotating_TOBA_Xbar}
\end{eqnarray}
where $I$ is the inertia moment of the bar and $\alpha$ is the
shape factor.
On the other hand, in the case of the thin bar aligned along $Y$
axis, we have $q^{XX}=q^{YY}=0$ and $q^{XY}=:-I\alpha$ and the
torque affects this thin bar is given by   
\begin{eqnarray}
  F_{gw(ybar)}
  &=&
  + \frac{I\alpha}{2}
  \left(
    \cos(2\Omega T)\ddot{h}_{\times}
    +
    \sin(2\Omega T)\ddot{h}_{+}
  \right)
  .
  \label{eq:Forces_affects_the_rotating_TOBA_Ybar}
\end{eqnarray}
In an approximation where the test-mass bar freely rotates
around the $Z$ axis, the equation of motion for the resultant
output of the antenna $\theta_{diff}$ is given by
\begin{eqnarray}
  I \ddot{\theta}_{diff}
  &=&
  F_{gw(xbar)}
  -
  F_{gw(ybar)}
  \nonumber\\
  &=&
  I\alpha
  \left(\cos(2\Omega T)\ddot{h}_{\times}
    +
    \sin(2\Omega T)\ddot{h}_{+}
  \right)
  .
  \label{eq:final-torque-equation}
\end{eqnarray}
This is completely identical to the result
(\ref{eq:resultant-output-Ando}) derived in
Ref.~\cite{M.Ando-etal-2010}.
In the paper~\cite{M.Ando-etal-2010},
Eq.~(\ref{eq:final-torque-equation}) was derived in the case
$\omega_{gw}\ll\Omega$ as mentioned in the above.
However, the derivation in this paper shows that the
limitation $\omega_{gw}\ll\Omega$ is not necessary and
Eq.~(\ref{eq:final-torque-equation}) is valid even in the case
$\omega_{gw}\simeq\Omega$.
Of course, the proper reference frame is valid only near the
rotation axis and Eq.~(\ref{eq:final-torque-equation}) is valid
only when the size $s$ of the antenna satisfies
$\omega_{gw}s,\Omega s\ll 1$.
If this limitation become serious, we have to evaluate the
next-order expression discussed by Li and
Ni~\cite{W.-Q.Li-W.-T.Ni-1979}.


K.~N. deeply thanks to Professor Masa-Katsu Fujimoto for his
valuable discussions and encouragements.
This work was supported by JSPS KAKENHI Grant Number 24244031.






\begin{thebibliography}{99}
\bibitem{KNcomment-gauge} 
  This coordinate transformation is so called the first kind
  gauge-transformation and have nothing to do with the second
  kind gauge-transformation.
  The second kind gauge degree of freedom is completely fixed
  through the TT-gauge.
\end{thebibliography}
\end{document}